\documentclass[10pt,twocolumn]{revtex4}
\usepackage{graphicx}
\usepackage{amssymb}

\begin{document}

\title{Eigen modes for the problem of anomalous light transmission \\
through subwavelength holes}

\author{ \vspace*{-3mm} B.\ Sturman and E.\ Podivilov}
\affiliation{Institute of Automation and Electrometry, 630090
Novosibirsk, Russia}

\author{\vspace*{-3mm} M.\ Gorkunov}
\affiliation{\mbox{Institute of Crystallography, Russian Academy
of Sciences, Leninskii pr.~59, 117333 Moscow, Russia}}

\begin{abstract}

We show that the wide-spread concept of optical eigen modes in
lossless waveguide structures, which assumes the separation on
propagating and evanescent modes, fails in the case of
metal-dielectric structures, including photonic crystals. In
addition to these modes, there is a sequence of new eigen-states
with complex values of the propagation constant and non-vanishing
circulating energy flow. The whole eigen-problem ceases to be
hermitian because of changing sign of the optical dielectric
constant. The new anomalous modes are shown to be of prime
importance for the description of the anomalous light transmission
through subwavelength holes.

\end{abstract}


 \maketitle

The concept of eigenmodes lies at the center of any physical problem
for optical light-transmitting systems including waveguides and
photonic crystals. This concept seems to be clear and the general
properties of the eigen-values and eigen-functions seem to be fully
presented in numerous textbooks, see,
e.g.,~\cite{Yariv,Vassalo,Collin,PC}.

For photonic crystals (PC), determination of the band structures
-- the eigen-frequencies and eigen-functions versus the wavevector
-- is the most common. The eigen-frequencies are proven to be real
for any lossless PC, i.e., for any periodic distribution of real
dielectric optical permittivity $\varepsilon({\bf r})$. A close
analogy with the electronic band states is well
established~\cite{PC}.

Another formulation of the eigen-mode problem is typical of
waveguiding systems which are uniform along the propagation
coordinate~$z$: To find the admitted values of the propagation
constant $\beta$ entering the propagation factor $\exp(i\beta z)$
and the corresponding transverse (in $x$ and $y$) distributions of
the optical light fields.

It is known for the lossless dielectric systems ($\varepsilon
> 0$) that the admitted values of $\beta^2$ are real -- positive and
negative~\cite{Yariv,Vassalo,Collin}. The corresponding eigenmodes
are propagating and evanescent, respectively, and a close analogy
with quantum mechanics holds true. The same is valid for the
ideal-metal waveguiding systems, $\varepsilon \rightarrow \infty$.

A permanently growing interest to metal-based systems is usually
attributed to the excitation of surface
plasmons~\cite{Yariv,Ebbesen2} which allow the light constrain on
a subwavelength scale. The necessary conditions, $\varepsilon' =
{\rm Re}\, \varepsilon < -1$, $\varepsilon'' = {\rm Im}\,
\varepsilon \ll 1$, are fulfilled for many metals. The limit of
lossless metal ($\varepsilon' < 0$, $\varepsilon'' = 0$) is as
actual as that of lossless dielectric. Recent discovery of the
extraordinarily light transmission through subwavelength holes in
metal slabs~\cite{Ebbesen2,Ebbesen1} has strongly enhanced the
current interest. However, the main ingredients and parametric
dependences of this fundamental phenomenon remain poorly
understood; its description relies on numerical simulations and
simplified models~\cite{Porto,Treacy,Lalanne,Zakharian}.

The present notion of eigenmodes for lossless ($\varepsilon'' =
0$) metal-dielectric waveguide structures (1D and 2D PCs, arrays
of holes and single holes in metals, etc.) is surprisingly scarce.
It is widely believed that the values of $\beta^2$ remain real,
i.e. the separation into the propagating and evanescent modes and
the analogy with quantum mechanics, still hold true. Using
particular models for $\varepsilon(x,y)$ and Fourier expansions,
many authors calculate numerically only the real values
$\beta^2(\kappa_x,\kappa_y)$, see, e.g.,~\cite{Treacy,Tikhodeev}
and references therein. To the best of our knowledge, there are
only few publications where the authors mention, on the basis of
particular calculations, that some of values of $\beta^2$ cease to
be real~\cite{Sheng,Whittaker}.

The aim of this letter is ({\it i}) to show that the generally
accepted concept of eigenmodes is insufficient for the case of
lossless metal-dielectric waveguide structures, ({\it ii}) to
analyze the general properties of new eigenmodes which are neither
propagating nor evanescent, and ({\it iii}) to demonstrate that
these new modes are crucial for the description of the anomalous
light transmission through subwavelength holes. We will also pay
attention to the propagating modes in the subwavelength case.

To catch the essence of the eigen-mode problem, we consider the
case of a 1D PC with $\varepsilon$ being a real periodic function
of the coordinate~$x$. Restricting ourself to the TH polarization,
we have from Maxwell equations for the only nonzero component of
the magnetic field $H = H_y$:
\begin{equation}
\hat{L} H = \beta^2 H\;, \quad \hat{L} = \varepsilon \frac{d}{dx}
\frac{1}{\varepsilon}\frac{d}{dx} + \varepsilon k_0^2 \;,
\end{equation}
\noindent where $k_0 = 2\pi/\lambda$. This sets an eigen-mode
problem for the operator $\hat{L}$ with $\beta^2$ being the
eigen-value. The nonzero components of the light electric field
are expressed by $H$: $E_x = \beta H/\varepsilon k_0$, $E_z =
(i/\varepsilon k_0)\,dH/dx$.

The conclusion about reality of $\beta^2$ and analogy with quantum
mechanics is generally based on Hermitian character of the
differential operator $\hat{L}$, i.e., on the property $\langle
H_2|\hat{L}H_1\rangle = \langle H_1|\hat{L}H_2\rangle^*$, where
$\langle ..|..\rangle$ stands for the scalar product. The only
possibility to get this property is to define the scalar product
of two complex functions $H_1$ and $H_2$ as
\begin{equation}
\langle H_2|H_1\rangle = \int \varepsilon^{-1}(x)\;
H_2^*(x)\,H_1(x)\; dx \;.
\end{equation}
\noindent In the dielectric case, $\varepsilon(x) > 0$, this
definition meets all axioms of the scalar product~\cite{Mathews}.
For sign-changing functions $\varepsilon(x)$, it contradicts,
however, to one of the axioms -- the norm $\langle H|H\rangle$
ceases to be positively defined. Therefore, the conclusions about
Hermitian character of $\hat{L}$ and about reality of $\beta^2$
cannot be made.

Apart from the above negative observation, some positive
assertions can be made: \\
-- If an eigen-function possesses a positive norm, the
corresponding value of $\beta^2$ is real. Two eigen-functions
$H_{1}$ and $H_{2}$ corresponding to different real eigenvalues
$\beta_1^2$ and $\beta_2^2$ are orthogonal, $\langle
H_2|H_1\rangle = 0$. If at least one of $\beta_{1,2}$ is complex,
it has to be replaced by $\langle H^*_2|H_1\rangle = 0$. \\
-- In the case of TE polarization (nonzero $E_y$, $H_x$, and $H_z$
components) the above peculiarities are absent -- the admitted
values $\beta^2$ are real and an analogy with quantum mechanics
holds true. \\
-- In the 2D case, Hermitian character of the eigen-mode problem
and reality of $\beta^2$ can be proven in exceptional cases of
waves with $E_z = 0$ (TE modes). In the general case (TM and
hybrid modes) complex values of $\beta^2$, i.e., anomalous modes,
are present, see also below.

Now we turn to particular examples. Consider first a 1D PC
consisting of an alternating sequence of layers with
permittivities $\varepsilon_{1,2}$ and thicknesses $x_{1,2}$; the
period is $x_0 = x_1 + x_2$. Solution of Eq.~(1) for an eigenmode
has the form $H(x) = e^{\displaystyle i\kappa x}\, h(x)$, where
$\kappa$ is the Bloch wavevector ranging from $-\pi/x_0$ to
$\pi/x_0$ and $h(x)$ is an $x_0$-periodic function. Using
continuity of $H(x)$ and $E_z(x) \propto \varepsilon^{-1}\,dH/dx$
at the interfaces and periodicity of $h(x)$ and $dh/dx$, we obtain
the following dispersion equation for $\beta^2$ (it generalizes
Eq.~(13) of~\cite{Sheng}):
\begin{eqnarray}\label{Disp}
\frac{1}{2}\left(\frac{p_1\varepsilon_2}{p_2\varepsilon_1} +
\frac{p_2\varepsilon_1}{p_1\varepsilon_2}
\right)\sin(p_1x_1)\,\sin(p_2x_2) \nonumber \\
= \cos(p_1x_1)\,\cos(p_2x_2) - \cos(\kappa x_0) \;,
\end{eqnarray}
\noindent where $p_{1,2} = (\varepsilon_{1,2}k_0^2 -
\beta^2)^{1/2}$. The trigonometric functions possess generally
complex arguments. If $\varepsilon_{1,2}$ are real, Eq.~(3) is
real as well. The roots $\beta = \pm \sqrt{\beta^2}$ describe the
wave propagation in the $\pm z$ directions.

The circles in Fig.~1a show the values of the propagation constant
on the complex $\beta',\beta''$ plane in the lossless case for
$\kappa = 0$ (at the center of the Brillouin zone), $x_1 = x_0/5 =
\lambda/4$, $\varepsilon_1 = 1$, and $\varepsilon_2 = -9.6$. The
last parameter corresponds to silver at $\lambda \approx
0.5\;\mu$m~\cite{Christy}.
\begin{figure}[h]
\centering\includegraphics[width=\columnwidth]{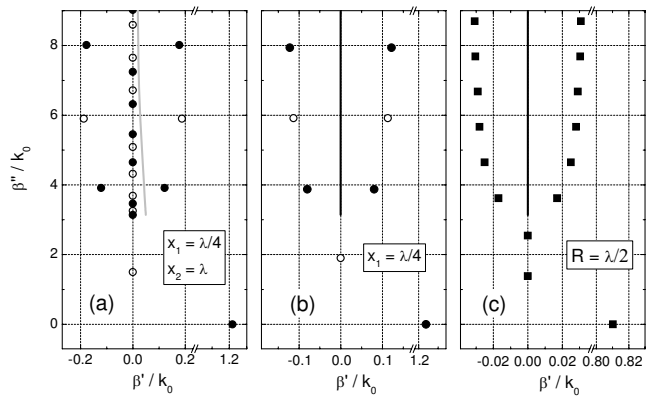}
\vspace{-0.2cm} \caption{Values of $\beta$ for $\varepsilon_2 =
-9.6$ and three structures: \hspace{5mm} (a) 1D PC with $x_2 = 4x_1$
(the gray line shows positions of evanescent roots for
$\varepsilon''_2 = 0.3$); (b) a single slit with $x_1 = \lambda/4$;
(c) a single cylindric hole with $R = \lambda/2$. The vertical solid
lines in (b) and (c) show the continuous spectrum.}\label{Fig.1}
\end{figure}
We have a single real root, $\beta/k_0 \approx 1.2$, and two
sequences of roots with~$\beta'' \neq 0$. One sequence, with pure
imaginary values of $\beta$, refers to the known evanescent modes.
Another sequence, with $\beta' \neq 0$ and $\beta'' \gtrsim
\sqrt{|\varepsilon_2|}$ (complex values of $\beta^2$) corresponds
to {\it anomalous modes} forbidden in the dielectric case. They
combine features of propagating and evanescent modes. The
symmetric pairs of anomalous roots correspond to mutually
conjugate values of $\beta^2$.

At $\kappa = 0$, the modes are either even, $h(x) = h(-x)$, or
odd, $h(x) = -h(-x)$; the "even" and "odd" roots are shown by
filled and open circles in Fig.~1a. The single real root
corresponds to the even propagating mode. About $30\%$ of the
modes are anomalous.

With weak losses taken into account, the roots experience small
displacements. The strongest of them (shifts to the right) occur
for the evanescent roots. The gray line in Fig.~1a shows positions
of these roots for $\varepsilon''_2 = 0.3$, which models silver at
$\approx 0.5\;\mu$m. The other displacements do not exceed the
circle size.

Whereas the Pointing vector ${\bf P} = (P_x,P_z)$ is zero for the
evanescent modes, it is nonzero for the anomalous modes. Each
anomalous mode is responsible thus for circulation of the light
energy. Applying the orthogonality relation $\langle
H^*_2|H_1\rangle = 0$ to a conjugate pair of anomalous modes, we
come to the following relation for an anomalous eigen-function
$h_a(x)$:
\begin{equation}\label{ZeroNorm}
\int \varepsilon^{-1}(x)\;|h_a(x)|^2\;dx = 0 \;,
\end{equation}
\noindent where the integration occurs over a period of~$x_0$.
Since $P_z \propto \beta'\varepsilon^{-1}(x)\,|h(x)|^2
\exp(-2\beta''z)$, it shows that the $x$-averaged energy flux is
zero for the anomalous modes -- the inflow in air is compensated
by the outflow in metal. Eq.~(\ref{ZeroNorm}) can be interpreted
also as a zero norm for the eigen-vector $h_a(x)$. The
eigen-vectors $h(x)$ for the propagating and evanescent modes
possess positive norms.
\begin{figure}[h]
\centering
\includegraphics[width=0.9\columnwidth]{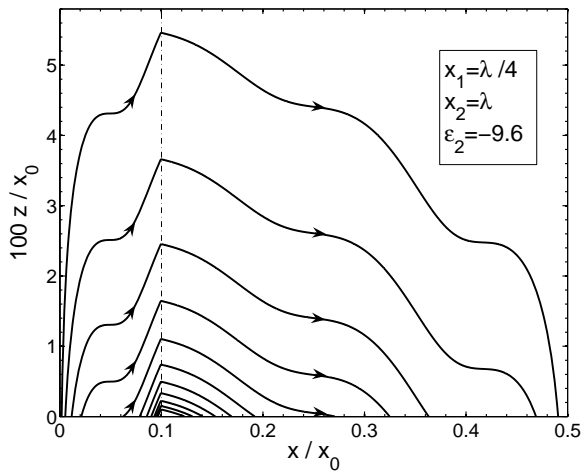}
\caption{Streamlines of the Pointing vector for the first anomalous
symmetric mode with $\beta' > 0$.}\label{Fig.2}
\end{figure}
\noindent Fig.~2 shows the energy circulation within a half-period
$x_0/2$ for the first anomalous mode with $\beta' > 0$. For the
mode with $\beta' < 0$ the arrows have to be inverted.

What happens with the above modes when changing $x_{1,2}$? With
decreasing slit width $x_1$, the propagating mode survives
($\beta'$ grows as $x_1^{-1}$), the evanescent mode with $\beta''
< \sqrt{|\varepsilon_2|}$ disappears, the other evanescent modes
experience minor changes, and the vertical distance between the
anomalous modes increases as $2\pi/x_1$.

Increasing wall parameter $x_2$ does not affect strongly the
propagating and anomalous modes in Fig.~1a. However, the density
of evanescent modes in the region $\beta'' >
\sqrt{|\varepsilon_2|}$ increases $\propto x_2$. The limit $x_2
\to \infty$ corresponds to the single-slit case. It is represented
by Fig.~1b. We have single localized (in $|x|$) propagating and
evanescent modes, a sequence of localized anomalous modes, and a
continuous spectrum of non-localized evanescent modes with $\beta'
= 0$ and $\beta'' > \sqrt{|\varepsilon_2|}$. This also follows
from a direct description of the single-slit case. The anomalous
modes represent thus properties of a single-hole.

For $k_0x_1 \ll 1$, $x_2 \to \infty$ (very narrow single slit) we
have for the single propagating mode and $\varepsilon''_2 \ll
|\varepsilon'_2|$:
\begin{equation}\label{SSL}
\beta \simeq \frac{1}{x_1}\;\bigg[\ln\bigg(\frac{|\varepsilon'_2|
+ 1}{|\varepsilon'_2| - 1}\bigg) +
\frac{2i\,\varepsilon''_2}{|\varepsilon'_2|^2 - 1} \; \bigg] \,.
\end{equation}
\noindent The effective refractive index $\beta'/k_0$ tends to
infinity for $x_1 \to 0$. In the ideal-metal limit
($|\varepsilon_2| \to \infty$) we have $\beta = 0$. For $x_{1,2}
\to \infty$ the propagating mode transforms to the usual surface
plasmon.

The eigen-functions for the above periodic and single-slit cases
(localized and delocalized) can be constructed readily from the
exponential functions $\exp(ip_{1,2}x)$ with proper values of
$p_{1,2}(\beta^2)$.

The Bloch dependences $\beta'(\kappa)$ and $\beta''(\kappa)$ for
the propagating and evanescent modes, which are present in the PC
literature, are insufficient in the metal-dielectric case where
the anomalous modes are present. Moreover, the anomalous modes
cause strong interaction of different branches and peculiarities
of Bloch diagrams.
\begin{figure}[h]
\centering \includegraphics[width=\columnwidth]{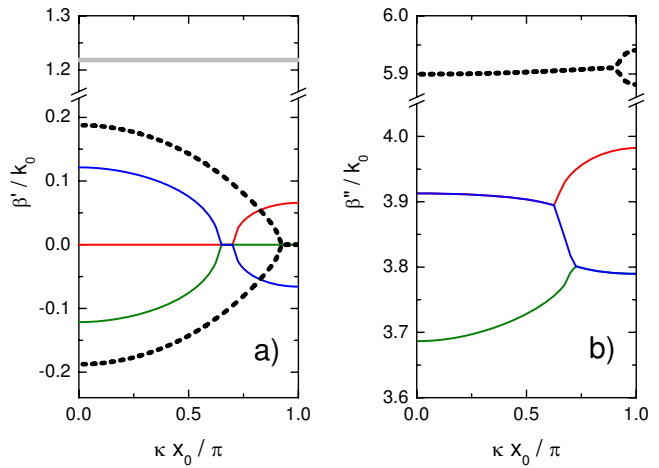}
\caption{Dependences $\beta''(\kappa)$ and $\beta'(\kappa)$ for the
periodic structure with $x_0 = 5x_1 = 5\lambda/4$, and
$\varepsilon_2 = -9.6$. Solid lines are plotted for the first pair
of anomalous modes and the nearest evanescent mode, the dotted lines
correspond to the second anomalous pair, and the gray line is for
the propagating mode.}\label{Fig.3}
\end{figure}
Fig.~3 shows examples of the $\kappa$-dependences for our 1D PC
structure. The value of $|\beta'|$ for the first pair of anomalous
modes (see Fig.~1a) decreases with growing $\kappa$ and turns to
zero for $0.66 \lesssim \kappa x_0/\pi \lesssim 0.71$, i.e., the
anomalous modes become evanescent within this interval. For
$\kappa x_0/\pi \gtrsim 0.71$ a pair of anomalous modes appears
again. This behavior is correlated with bifurcations of
$\beta''(\kappa)$ for the anomalous modes and the nearest
evanescent mode. Behavior of the second pair of anomalous modes is
different. The confluence of $\beta'(\kappa)$ for these modes at
$\kappa x_0/\pi \simeq 0.9$ is accompanied by the split of
$\beta''(\kappa)$. The dependence $\beta(\kappa)$ for the
propagating mode and most of evanescent modes is fairly weak.

The situation described is also inherent in 2D structures.
Consider for simplicity TM ($H_z = 0$) modes for a cylindric hole
of radius $R$ in a lossless metal. The corresponding dispersion
equation reads~\cite{Pfeiffer}
\begin{equation}\label{Bessel}
p_2\,J_1\,H^{(1)}_0 - \varepsilon_2p_1\,J_0\,H^{(1)}_1 = 0 \;,
\end{equation}
\noindent where $J_{0,1} = J_{0,1}(p_1R)$ and $H^{(1)}_{0,1} =
H^{(1)}_{0,1}(p_2R)$ are the Bessel and Hankel functions. Fig.~1c
shows the spectrum of $\beta$ for $\varepsilon_2 = -9.6$ and $R =
\lambda/2$. One propagating mode, two evanescent modes, a sequence
of localized anomalous modes, and a continuous spectrum are
present.

The TM propagating mode exists here only for $R >
R_{min}(\varepsilon_2)$ with $R_{min}$ decreasing from $\simeq
0.334\lambda$ to $\simeq 0.257\lambda$ when $\varepsilon_2$
changes from $-10$ to $-1$. For the HE$_{11}$ propagating mode,
which provides the minimum cut-off, $R_{min}$ decreases from
$\simeq 0.22\lambda$ to $\simeq 0.087\lambda$ in this range of
$\varepsilon_2$. For the ideal metal $R_{min}/\lambda \simeq 0.3$.
As follows from Eq.~(\ref{Bessel}), in the range $-1 <
\varepsilon_2 < 0$ a propagating mode exists even for $R \to 0$.

To show the importance and applications of our approach, we
consider the following fundamental problem: An $x$-polarized plane
wave of a unit amplitude is incident normally onto the above 1D
metal-dielectric PC occupying the semi-space $z > 0$. It is
necessary to determine the transmission properties at the
interface $z = 0$.

We use the following explicit mode expansion for $H$:
\begin{equation}
H = \left\{
\begin{array}{ll}
\sum\limits_{\nu = 0}^{\infty} \; a_{\nu} h^{\nu}(x)
\;e^{\displaystyle i\beta_{\nu}z}\;; & z > 0 \\
\phantom{\sum_{\nu = 0}^{\infty} \; a_{\nu} h^{\nu}(x)}
\\
e^{\displaystyle ik_0z} + \sum\limits_{n = -\infty}^{\infty} \;
b_n\,e^{\displaystyle inKx + q_n z} \;; & z < 0 \;,
\end{array}
\right.
\end{equation}
\noindent where $\nu$ numerates the even eigen-modes at $\kappa =
0$, $K = 2\pi/x_0$, $q_n = (n^2K^2 - k_0^2)^{1/2}$ for $n^2K^2 >
k_0^2$, $q_n = -i\,(k_0^2 - n^2K^2)^{1/2}$ for $k_0^2 > n^2K^2$,
and $b_n$ and $a_{\nu}$ are the amplitudes to be found. Real and
imaginary values of $q_n$ correspond to evanescent and propagating
waves in air. With increasing period, new diffraction orders
appear at $x_0 = n\lambda$ in reflected light; only a single
propagating wave (with $n = 0$) is allowed for $x_0 < \lambda$.

Employing the boundary conditions at the interface $z = 0$ and the
Fourier transformation, we come to the set of linear algebraic
equations for determination of $a_{\nu}$:
\begin{equation}\label{a-Equation}
\sum_{\nu = 0}^{\infty} \; a_{\nu} \left[\, h^{\nu}_n - i\,
\beta_{\nu} q_n^{-1}\; (h^{\nu}/\varepsilon)_n \, \right] =
2\delta_{n0} \;,
\end{equation}
\noindent where $h^{\nu}_n$ and $(h^{\nu}/\varepsilon)_n$ are the
Fourier components of the periodic functions $h^{\nu}(x)$ and
$h^{\nu}(x)/\varepsilon(x)$. These components were expressed
explicitly through $\beta_{\nu}$.

A numerical routine with truncation of this set at $\nu_{max} =
n_{max} = N \gg 1$ was used to calculate $a_{\nu}$ and determine
the reflection and transmission properties. The results of
calculations converge perfectly well with increasing $N$ when all
eigen-modes are taken into account. However, the calculation
results {\it diverge when the anomalous modes are ignored}. This
means (i) that the set of the eigen-functions is incomplete
without the anomalous modes and (ii) that these modes are directly
involved in the energy transfer from the incident plane wave to
the propagating mode(s) inside the slits. \vspace*{2mm}

\begin{figure}
\centering \includegraphics[width=0.9\columnwidth]{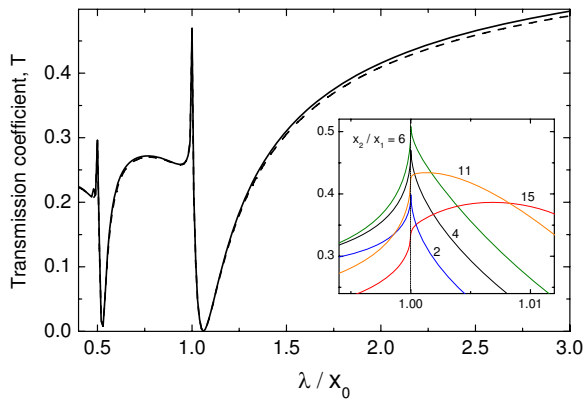}
\caption{Dependence $T(\lambda/x_0)$ for $\varepsilon'_2 = - 9.6$,
$x_2/x_1 = 4$, and $N = 48$; the solid and dashed lines are plotted
for $\varepsilon''_2 = 0$ and $0.3$. The inset shows
$T(\lambda/x_0)$ for $\varepsilon''_2 = 0$ and five values of
$x_2/x_1$ in the vicinity of the main peak.}\label{Fig.4}
\end{figure}

Consider the key results for the transmission coefficient $T$
defined as the energy-flax fraction transmitted into the
propagating mode at $z = 0$. This definition is useful even for
$\varepsilon'' \neq 0$ when the propagating mode decays inside the
metal as $\exp(-2\beta''z)$. Fig.~4 shows the dependence
$T(\lambda)$ for the lossless and lossy cases. The losses
influence $T$ very weakly; their main effect occurs during
propagation. The function $T(\lambda)$ possesses narrow peaks at
$\lambda = x_0/n$; the main peak corresponds to $n = 1$. The
geometric nature of the resonances is evident and it is not
overshadowed by the Fabry-Perot interference typical of the slab
case~\cite{Porto}. The maximum value of the transmission
coefficient, $T_{max} \simeq 0.47$ (which corresponds to $x_1 =
\lambda/5$) is pretty high. It is decreasing slowly with
increasing $|\varepsilon_2|$. For $\lambda
> x_0$, the difference $1 - T$ is the reflection coefficient.

Our procedure allows to resolve the fine structure of the main
peak and investigate the dependence $T_{max}$ on the ratio
$x_2/x_1$, see the inset in Fig.~4. The structure is different for
$x_2/x_1 < |\varepsilon_2|$ and $x_2/x_1 > |\varepsilon_2|$. In
the first case, the peak spike is sharp and centered exactly at
$\lambda = x_0$. In the second case, the right wing of the peak is
broadened and the very maximum is slightly shifted to the right.
Surprisingly, $T_{max}(x_1/x_2)$ {\it is not} a monotonously
increasing function; the optimum value of $x_1/x_2$ is $\simeq
1/6$ where $T_{max} \simeq 0.51$. Note that the effective
permittivity of our medium in the limit $\lambda \gg x_0$ is
$\langle \varepsilon^{-1} \rangle^{-1} = |\varepsilon_2|x_0/
(x_1|\varepsilon_2| - x_2)$, see, e.g.,~\cite{JOSA77}; it turns to
infinity and changes sign at $x_2/x_1 = |\varepsilon_2|$. Validity
of this limiting case was controlled numerically.

In conclusion, the eigen-mode problem for metal-dielectric
structures -- photonic crystals, single holes in a metal -- is
essentially different from that typical of dielectric structures
and quantum mechanics. It is not hermitian and the eigen-values
for the localized states are generally complex in the lossless
case. The anomalous transmission through subwavelength holes is
substantially determined by the transmission coefficient for a
single interface. This characteristic possesses a number of
fundamental features. The opposite faces of the metal-dielectric
structure are coupled via a weakly decaying propagating mode.

\end{document}